\documentclass[10pt,a4paper]{scrartcl} % KOMA-Script article scrartcl
\usepackage{palatino}
\PassOptionsToPackage{utf8}{inputenc}
	\usepackage{inputenc}

\PassOptionsToPackage{eulerchapternumbers,listings,%
					 pdfspacing,%floatperchapter,%linedheaders,%
					 subfig,beramono,eulermath,nochapters}{classicthesis}                                        
\PassOptionsToPackage{american}{babel}   % change this to your language(s)
	\usepackage{babel}                  

\usepackage{csquotes}
\PassOptionsToPackage{%
    backend=bibtex,bibencoding=ascii,%
    language=auto,%
    style=numeric-comp,%
    sorting=nyt, % name, year, title
    maxbibnames=10, % default: 3, et al.
    natbib=true % natbib compatibility mode (\citep and \citet still work)
}{biblatex}
    \usepackage{biblatex}

\PassOptionsToPackage{fleqn}{amsmath}       % math environments and more by the AMS 
    \usepackage{amsmath}

\PassOptionsToPackage{T1}{fontenc} % T2A for cyrillics
    \usepackage{fontenc}     
\usepackage{textcomp} % fix warning with missing font shapes
\usepackage{scrhack} % fix warnings when using KOMA with listings package          
\usepackage{xspace} % to get the spacing after macros right  
\usepackage{mparhack} % get marginpar right
\usepackage{fixltx2e} % fixes some LaTeX stuff --> since 2015 in the LaTeX kernel (see below)
\PassOptionsToPackage{printonlyused,smaller}{acronym} 
    \usepackage{acronym} % nice macros for handling all acronyms in the thesis
\usepackage{tabularx} % better tables
\usepackage{caption}
\usepackage{subfig}  
\usepackage{listings} 
\PassOptionsToPackage{pdftex,hyperfootnotes=false,pdfpagelabels}{hyperref}
    \usepackage{hyperref}  % backref linktocpage pagebackref
\PassOptionsToPackage{pdftex}{graphicx}
    \usepackage{graphicx}

\hypersetup{%
    colorlinks=true, linktocpage=true, pdfstartpage=3, pdfstartview=FitV,%
    breaklinks=true, pdfpagemode=UseNone, pageanchor=true, pdfpagemode=UseOutlines,%
    plainpages=false, bookmarksnumbered, bookmarksopen=true, bookmarksopenlevel=1,%
    hypertexnames=true, pdfhighlight=/O,%nesting=true,%frenchlinks,%
    urlcolor=webbrown, linkcolor=RoyalBlue, citecolor=webgreen, %pagecolor=RoyalBlue,%
}   

\usepackage{classicthesis} 
\usepackage{tabulary}
\usepackage{booktabs}
\usepackage[left=1in,right=1in,top=1in,bottom=1in]{geometry}

\bibliography{arxiv-biblio}

\newcommand{\ie}{\emph{i.\,e.}}

\newcommand{\eg}{\emph{e.\,g.}}
 
\newcommand{\wrt}{with respect to } 
\newcommand{\spara}[1]{\smallskip\noindent{\bf #1.}}

\title{\rmfamily\normalfont\spacedallcaps{Women Through the Glass Ceiling: Gender Asymmetries in Wikipedia}}
\author{Claudia Wagner\thanks{GESIS - Leibniz Institute for the Social Sciences, Cologne, Germany, and University of Koblenz-Landau, Koblenz, Germany.} 
\and Eduardo Graells-Garrido\thanks{Telefónica I+D, Santiago, Chile.} 
\and David Garcia\thanks{ETH Zurich, Zurich, Switzerland.}
\and Filippo Menczer\thanks{Center for Complex Networks and Systems Research, School of Informatics and Computing, Indiana University, USA.}}

\date{}

\begin{document}

\maketitle

\begin{abstract} 

Contributing to the writing of history has never been as easy as it is today thanks to Wikipedia, a community-created encyclopedia that aims to document the world's knowledge from a neutral point of view.
Though everyone can participate it is well known that the editor community has a narrow diversity, with a majority of white male editors.
While this participatory \emph{gender gap} has been studied extensively in the literature, this work sets out to \emph{assess potential gender inequalities in Wikipedia articles} along different dimensions: notability, topical focus, linguistic bias, structural properties, and meta-data presentation. 

We find that 
(i)~women in Wikipedia are more notable than men, which we interpret as the outcome of a subtle glass ceiling effect;
(ii)~family-, gender-, and relationship-related topics are more present in biographies about women; (iii)~linguistic bias manifests in Wikipedia since abstract terms tend to be used to describe positive aspects in the biographies of men and negative aspects in the biographies of women; 
and (iv)~there are structural differences in terms of meta-data and hyperlinks, which have consequences for information-seeking activities.
While some differences are expected, due to historical and social contexts, other differences are attributable to Wikipedia editors. The implications of such differences are discussed having Wikipedia contribution policies in mind.
We hope that the present work will contribute to increased awareness about, first, gender issues in the content of Wikipedia, and second, the different levels on which gender biases can manifest on the Web.

\spara{Keywords}
Wikipedia;
Gender Inequality;
Historical Relevance;
Lexical Bias;
Linguistic Bias;
Network Structure.

\end{abstract}

\section{Introduction}
\label{sec:intro}

Wikipedia aims to provide a platform to freely share the sum of all human knowledge. It represents an influential source of information on the Web, containing encyclopedic information about notable people from different countries, epochs, and disciplines.
It is also a community-created effort driven by a self-selected set of editors.  
In theory, by following its guidelines about verifiability, notability, and neutral point of view, Wikipedia should be an unbiased source of knowledge. In practice, the community of Wikipedians is not diverse, but predominately white and male \cite{lam2011wp,Collier2012,hill2013wikipedia}, and women are not being treated as equals in the community \cite{lam2011wp}. 
In our previous work we found that gender asymmetries exist in Wikipedia content \cite{wagner2015s,graells2015first}. Here we extend our prior work and provide an in-depth analysis of who makes it into Wikipedia and how these people are presented. 

\textbf{Objectives:} 
This work sets out to \emph{assess potential gender inequalities in Wikipedia articles} along different dimensions. Concretely, we aim to address the following research questions:
(i)~Are men and women who are depicted in Wikipedia equally notable --- \ie, do Wikipedians use the same thresholds for women and men when deciding who should be depicted on Wikipedia? 
(ii)~Are any topical aspects overrepresented in articles about men or women?
(iii)~Does linguistic bias manifest in Wikipedia? 
(iv)~Do articles about men and women have similar structural properties, \ie, similar meta-data, and network properties in the hyperlink network? 

\textbf{Approach:} 
We define gender inequality as a \emph{systematic asymmetry} \cite{beukeboom2014mechanisms} in the way that the two genders are treated and presented. 
To assess the extent to which Wikipedia suffers from potential gender bias, we compare biographies about men and women in Wikipedia along the following dimensions: external and internal global notability, topical and linguistic presentation, structural position, and meta-data presentation.

\textbf{Contributions \& Findings:} 
Our results show that:
\begin{itemize}

\item Women in Wikipedia are on average slightly more notable than their male counterparts. Furthermore, the gap between the number of men and women is larger for ``local heroes'' (people who are only depicted in few language editions) than for ``superstars'' (people who are present in almost all language editions). These effects can be explained by interpreting Wikipedia's entry barrier as a subtle \emph{glass ceiling.} While it is obvious that very notable people should be included in Wikipedia, the decision is questionable for people who are less notable. We find that bias and inequality manifest themselves in the presence of such uncertainty, as the Wikipedia editor community must make more subjective decisions about inclusion.

\item There are differences in the topical focus of biographical content, where gender-, family-, and relationship-related topics are more dominant in the stand-alone overviews of biographies about women in the English Wikipedia.

\item Linguistic bias becomes evident when looking at the abstractness and positivity of language. Abstract terms tend to be used to describe positive aspects in biographies of men, and negative aspects in biographies of women. 

\item There are structural differences in terms of meta-data and hyperlinks, which have consequences for information-seeking activities.

\end{itemize}
The contributions of this work are twofold: (i)~we present a computational method for assessing gender bias in Wikipedia \emph{along multiple dimensions} and (ii)~we apply this method to the English Wikipedia and share empirical insights on the observed gender inequalities. 
The methods presented in this paper can be used to assess, monitor and evaluate these issues in Wikipedia on an ongoing basis. 
We translate our findings into potential actions for the Wikipedia editor community to reduce gender bias in the future.

\section{Data \& Methods}
\subsection{Dataset}
\label{sec:dataset}

To study gender bias in Wikipedia, we consider the following data sources:

\begin{enumerate}
 \item The DBpedia 2014 dataset \cite{dbpediaswj}.\footnote{\url{http://oldwiki.dbpedia.org/Downloads2014}}
 \item Inferred gender for Wikipedia biographies by \cite{bamman2014unsupervised}.\footnote{\url{http://www.ark.cs.cmu.edu/bio/}}
\end{enumerate}

DBpedia \cite{dbpediaswj} is a structured version of Wikipedia that provides meta-data for articles; normalized article \emph{Uniform Resource Identifiers} (URIs) that allow to interlink articles about the same entity in different language editions; normalized links between articles (taking care of redirections); and a categorization of articles into a shallow ontology, which includes a \emph{Person} category.
This information is available for 125 Wikipedia editions. 

To obtain gender meta-data for biographies in the English Wikipedia edition we match article URIs with the dataset by Bamman and Smith \cite{bamman2014unsupervised}, which contains inferred gender for biographies based on the number of grammatically gendered words (\eg, \emph{he}, \emph{she}, \emph{him}, \emph{her}, etc.). 
Note that only \textit{male} and \textit{female} genders are considered in this dataset. 
The gender meta-data in other language editions are obtained from Wikidata by exploiting the links between DBpedia and Wikidata.
Wikidata reports more genders (\eg, \textit{transgender male} and \textit{transgender female}). However, those genders have a very small presence, and thus we only focus on \textit{male} and \textit{female}.

Table~\ref{table:dataset} shows the biography statistics of the 20 largest Wikipedia editions in terms of entities available with meta-data in DBpedia.
The English edition contains the largest number of biographies with gender information (893,380), while the Basque edition (eu) contains the lowest number of biographies (3,449).
In terms of representation of women, 15.5\% of biographies in the English edition are about women. The smallest fraction of women can be found in the German edition (13.2\%), while the maximum fraction is found in the Korean edition (22.6\%).
Since the English language edition has the largest number of articles covering personalities from multiple editions and all language editions share in average 97\% of people with the English language editions, we focus our analysis on the English edition.

We split this dataset in Pre-1900 and Post-1900. The Pre-1900 sample contains all people born before 1900, while the Post-1900 sample consists of people born in or after 1900.

\subsection{Approach}
\label{sec:approach}
To assess the extent to which gender bias manifests in Wikipedia, we compare Wikipedia articles about men and women along the following dimensions: 

\begin{enumerate}
 \item Global notability of people according to external and internal proxy measures.
 \item Topical focus and linguistic bias of biography articles.
 \item Structural properties of articles, including meta-data and network-theoretic position of people in the Wikipedia article link network.
\end{enumerate}

\subsubsection{Global Notability}
Let us first compare how difficult it is for men and women to make it into Wikipedia. 
Do Wikipedians use the same notabilfity threshold for men and women when deciding who should be included? Or does the so called \emph{glass-ceiling effect} make it more difficult for women to be recognized for their achievements? 
Recall that the glass-ceiling effect refers to the situation in which women cannot reach higher positions because an ``invisible barrier'' (namely, gender bias) prevents them from doing so. 

We hypothesize that if the entry point of Wikipedia functions as a glass ceiling, fewer women will be included in Wikipedia, but those women will be more notable than their male counterparts on average.
Especially if we compare the number of male and female ``local heroes'' (people with low levels of notability, without worldwide fame), we expect to see a larger gender gap (\ie, fewer women than men) than for worldwide ``superstars,'' because fewer female ``local heroes'' will be able to make it into Wikipedia. 

To address the question of whether a glass-ceiling effect exists in Wikipedia, we study the population of men and women who are depicted in Wikipedia and analyze their global notability from an \emph{internal and external perspective}. 

Assessing the notability of people is a difficult task. Fortunately, Wikipedia and search engines like Google allow us to gauge public interest in different people and from different locations over time. Such signals can be employed as proxies for the notability of people. These proxy measures are noisy and may also be biased, since they reflect the interests of Google users or Wikipedia editors, which in turn are influenced by many factors. 
Nevertheless, both signals that we explore let us compare the public interest in men and women. 
While our analysis allows us to quantify the existence of a glass-ceiling effect, it does not permit an assessment of its origin. It could be that Wikipedians unconsciously apply different thresholds for men and women or that Wikipedia only reflects the glass ceiling of our society and other media, which only document the life of women who have higher capacities and abilities than men which are covered.

Concretely, we use the following external and internal proxy measures:

\smallskip
\textbf{Number of Language Editions:} The number of Wikipedia language editions that contain an article about a person is used as an internal proxy measure for that person's global notability. The idea is that people who only show up in a few language editions are less relevant from a global perspective than those who show up in more language editions. 
The DBpedia dataset provides a mapping for articles between different language editions, enabling us to count the number of editions in which a biography appears. 
In particular, we consider the biographies that appear in at least one of the top 20 languages of DBpedia, and count how often they show up in any other language editions. 

To explore whether the number of editions is influenced by gender, we fit a \textit{negative binomial} (NB) regression model. The number of editions in which a person is depicted is used as dependent variable, while gender is used as independent variable. We include the profession of a person (obtained through the DBpedia ontology classes) as well as the decade in which the person was born (obtained from the DBpedia date of birth meta-data) as control variables.
The NB model is appropriate since we consider overdispersed count data.

\smallskip
\textbf{Google Search Volume:} The Google trend\footnote{\url{https://www.google.com/trends/}} data gauge the interest of Google users between 2004 and 2015. Google trend data serve as an external proxy for the public interest toward a person, or information need about that person, and can be measured in different countries and at different points in time.

For a random sample of around 5000 people born after 1900 and before 2000 we collected Google trend data using the full name of the person as input. Google trends shows how often search terms are entered in Google relative to the total search volume in a region or globally.
Using full names as search terms will of course introduce noise since several people may share the same name. However, a similar level of noise can be expected for men and women.

We count the number of countries and the number of months between January 2004 and October 2015 (from a worldwide perspective) that reveal a relative search volume above a threshold which is chosen by Google. 
The Google threshold is relative to the total number of searches in the region and month under consideration.
To explore whether the number of countries and number of months in which we observe search volume above the threshold is influenced by gender, we fit two linear regression models that both use gender as the independent variable.
We also used a negative binomial regression model and obtained similar results, but a loss of power.

\subsubsection{Topical and Linguistic Bias}
After the investigation of potential differences in entry barriers, let us focus on the lexical presentation of those who made it into Wikipedia. 
Language use is reportedly different when speaking about different genders \cite{lakoff1972language}.
For example, the \emph{Finkbeiner test} \cite{finkbeinertest} suggests that an article about a woman often emphasizes the fact that she is a woman, mentions her husband and his job, her children and childcare arrangements, how she nurtures her underlings, how she is taken aback by the competitiveness in her field, and how she is such a role model for other women. 
Historian Gillian Thomas investigated the role of women in Encyclopaedia Britannica, finding that as contributors, women were relegated to matters of ``social and purely feminine affairs'' and as subjects, women were often little more than addenda to male biographies (\eg, Marie Curie as the wife of Pierre Curie) \cite{Thomas1992}.

Beside topical bias, previous research also suggests that linguistic biases may manifest when people describe other people that are part of their in- or out-group~\cite{Otterbacher2015}.
Linguistic bias is a systematic asymmetry in language patterns as a function of the social group of the persons described, and is often subtle and therefore unnoticed.
The Linguistic Intergroup Bias (LIB) theory \cite{Maass1989} suggests that for members of our in-group, we tend to describe positive actions and attributes using more abstract language, and their
undesirable behaviors and attributes more concretely. In other words, we generalize their success but not their failures. 
Note that verbs are usually used to make more concrete statements (\eg, ``he failed in this play''), while adjectives are often used in abstract statement (\eg, ``he is a bad actor'').
Conversely, when an out-group individual does or is something desirable, we tend to describe them with more concrete language (we do not generalize their success), whereas their undesirable attributes are encoded more abstractly (we generalize them).
Maass~\textit{et al.} point out that LIB may serve as a device that signals to others both our status
with respect to an in- or out-group, as well as our expectations for their behavior and attributes \cite{Maass1989}.
Our expectations are of course not only determined by our group-membership but also by the society in which we live. For example, in some situations or domains not only men but also women may expect other women to be inferior to men.

While it is well known that topical and linguistic biases exist, it is unknown to what extent these biases manifest in Wikipedia.
To investigate this question we compare the overview of biographies about men and women in the English Wikipedia. The overview (also known as lead section) is the first section of an article. According to Wikipedia, it ``should stand on its own as a concise overview of the article's topic. It should define the topic, establish context, explain why the topic is notable, and summarize the most important points.''\footnote{\url{https://en.wikipedia.org/wiki/Wikipedia:Manual_of_Style/Lead_section}} 
We focus on the lead section for two reasons. On one hand, the first part of the article is potentially read by most people who look at the article. On the other hand, Wikipedia editors need to focus on what they consider most important about the person, and biases are likely to play a role in this selection process.

\smallskip
\textbf{Topical Bias:}
To unveil topical biases in Wikipedia content, we analyze the following three topics that could be over-represented in articles about women according to what is suggested by Thomas's observations in Britannica and the Finkbeiner test:
\begin{itemize}
\item The \emph{gender} topic contains words that emphasize that someone is a man or woman (\ie, man, women, mr, mrs, lady, gentleman) as well as sexual identity (\eg, gay, lesbian).
\item The \emph{relationship} topic consists of words about romantic relationships (\eg, married, divorced, couple, husband, wife).
\item The \emph{family} topic aggregates words about family relations (\eg, kids, children, mother, grandmother).
\end{itemize}

To associate words with these topics (plus an unrelated category, \textit{other}), we follow an open vocabulary approach \cite{schwartz2013personality}. 
Because we want to include concepts that may comprise more than one word, we consider n-grams with $n \leq 2$.
We then analyze the association between the top 200 n-grams for each gender and the four topics (gender, relationship, family, or other). 
To rank the n-grams for men and women we use \textit{Pointwise Mutual Information} \cite{church1990word}. PMI measures the relationship between the joint appearance of two outcomes (X and Y) and their independent appearances. It is defined as:
\[
\mbox{PMI}(X, Y) = \log \frac{P(X, Y)}{P(X) P(Y)}
\]
where, in our case, $X$ is a gender and $Y$ is an n-gram.
The value of $P(X)$ can be estimated from the proportions of biographies about men and women, and the other probabilities can be estimated from n-gram frequencies.
PMI is zero if $X$ is independent of $Y$, it is greater than 0 if $X$ is positively associated with $Y$, and it is smaller than 0 if $X$ is negatively associated with $Y$.
We exclude words that appear in biographies from one gender only, 
because such words have undefined PMI for the other gender, and thus the comparison is not meaningful.
We are interested in words/n-grams that may appear in any gender, and which presumably could be independent of gender.
Finally, we compare the proportion of topics that are present in the top 200 n-grams that we associated with men and women using chi-square tests. In the absence of topical asymmetries, one would expect to observe only minor differences in the proportions of topics for men and women.

\smallskip
\textbf{Linguistic Bias:}

To measure linguistic bias, we use a lexicon-based approach and syntactic
annotations to detect abstract and subjective language as proposed by Otterbacher
\cite{Otterbacher2015}. The level of abstraction of language can be detected
through the syntactic class of terms, where adjectives are the
most abstract class, as for example comparing ``is violent'' with ``hurt the
victims'' \cite{Gorham2006}. 

To test for the existence of linguistic biases in Wikipedia, we quantify the
tendency of expressing positive and negative aspects of biographies with
adjectives, as a measure of the degree of abstraction of positive and negative
content.  
We quantify the tendency to use abstract language in each class as
the ratio of adjectives among positive and negative words. To do so, we detect positive and negative terms taken from the
\emph{Subjectivity Lexicon} \cite{Wilson2005}. For each term that in the lexicon, we check if it is an adjective or not based on part-of-speech tags \cite{Bird2009}.

After processing the text, we count for each biography the numbers of positive
$W_{+}$ and negative $W_{-}$ words, and from those the numbers of positive
adjectives $A_{+}$ and negative adjectives $A_{-}$. We combine these
counts into ratios of abstract positivity and negativity computed as
$r_{+}=A_{+}/W_{+}$ and $r_{-}=A_{-}/W_{-}$. This way, we quantify the
tendency to generalize positive and negative aspects of the biographies, with
the purpose of testing if this generalization depends on the gender of the
person being described.

The presence of gender stereotypes and sexism and the Linguistic Intergroup Bias (LIB) theory suggest that abstract terms would be more likely to be used to describe positive
aspects in the biographies of men than in biographies of women. 
Similarly, abstract language would be more likely to describe negative aspects in the biographies of
women in comparison to biographies of men. 
We test this hypothesis first
through a chi-square test on the aggregated ratios of adjectives over positive
and negative words in all biographies of each gender. To test if the bias
appears at the individual level, we then focus on biographies with at least 250
words and one evaluative term, testing if the measured $r_{+}$ and $r_{-}$
depends on gender while controlling for professions and the century in which a person was born.

\subsubsection{Structural Properties}
Structural properties impact how visible and reachable articles about notable men and women are, since users and algorithms rely on this information when navigating  
Wikipedia or when assessing the relevance of content within a certain context. %, among other contexts. 
For instance, search result rankings are often informed by centrality measures such as PageRank. Furthermore, search results show meta-data when the query is related to notable personalities (using, \eg, the Google Knowledge Graph \cite{singhal2012introducing}). 
These examples show that gender inequalities that manifest in the structure of Wikipedia may have important implications since they impact the information consumption process. 
 
\smallskip
\textbf{Meta-data:}
To provide structured meta-data, DBpedia processes content from the infoboxes in Wikipedia articles.
The infoboxes are tables with specific attributes that depend on the main activity associated with the person portrayed in the article. For instance, anyone has attributes like date/place of birth, but philosophers have ``Main Ideas'' in their attributes, and soccer players have ``Current Team'' as an attribute.
To explore asymmetries between attribute distributions according to gender, we first identify all meta-data attributes present in the dataset. Then, for each attribute we count the number of biographies that contain it. Finally, we compare the relative proportions of attribute presence between genders using chi-square tests, considering the male proportion as baseline, and discuss which differences go beyond what can be explained by professional areas. 

\smallskip
\textbf{Hyperlink Network:}
We build a network of biographies using the hyperlink structure among Wikipedia articles about people in the English language edition. Concretely, we use the structured links between the canonical URLs of  articles provided by DBpedia, where redirects are resolved. 
On this network we perform two different analyses: first, we explore to what extent the connectivity between people is influenced by gender, and second, we investigate the relation between the centrality of people and their gender. 
To this end, we compute the PageRank of articles about people. PageRank is a widely used measure of network centrality %that is based on network connectivity 
\cite{brin1998anatomy,Fortunato08internetmath}. 
To explore potential asymmetries in network centrality, we sort the list of biographies according to their PageRank values in descending order. 
We estimate the fraction of biographies that are about women at different ranks $k$.
In the absence of any kinds of inequality, whether endogenous or exogenous to Wikipedia, one would expect the fraction of women to be around the overall proportion of women biographies, irrespective of $k$. 

To discern whether the observed asymmetries \wrt gender go beyond what we would expect to observe by chance, we compare our empirical results 
with those obtained from baseline graphs that are constructed as follows:
\begin{itemize}
 \item \textit{Random}. 
We shuffle the edges in the original network. 
For each edge (u,v), we select two random nodes (i,j) and replace (u,v) with (i,j). The resulting network is a random graph with neither the heterogeneous degree distribution nor the clustered structure that the Wikipedia graph reveals \cite{zlatic2006wikipedias}.

 \item \textit{Degree Sequence}. 
We generate a graph that preserves both in-degree and out-degree sequences (and therefore both distributions) by shuffling the structure of the original network. 
For a random pair of edges ((u,v), (i,j)) rewire to ((u,j), (i,v)). 
We repeat this shuffling as many times as there are edges. Note that although the in- and out-degree of each node are unchanged, the degree correlations and the clustering are lost.
 \item \textit{Small World}. 
We generate an undirected small world graph using the model by Watts and Strogatz 
\cite{watts1998collective}. 
This model interpolates a random graph and a lattice in a way that preserves two  properties of small world networks: average path length and clustering coefficient. After building the graph, we randomly assign a gender to each node, maintaining the proportions from the observed network. 
\end{itemize}

\subsection{Tools}

We provide implementations of our methods, as well as data-gathering tools, in a public repository available at \url{github.com/clauwag/WikipediaGenderInequality}.

\section{Results}

In this section we present the results of our empirical study about gender inequalities in Wikipedia.

\subsection{Inequalities in Global Notability Thresholds}

Let us first test our hypothesis that the Wikipedia entry point functions as a glass ceiling, making it more difficult for women to be included. If this is the case, women who made it into Wikipedia should be more notable than men. 
We measure notability using the internal and external proxies based on language editions and search volume, respectively.
We filtered biographies that did not have a birth date in their meta-data, as well as those with birth date previous to year 0, and those with birth date greater than year 2015. 
Consequently, in this analysis we consider $N=$ 590,741 biographies (with 14.7\% women). % before: 590,124
In addition to examining all biographies at once, we split the dataset in two parts 
to account for the fact that the visibility of women and presumably also their access to resources has changed drastically over time. 
We thus consider biographies of people born before 1900 ($N_b=$ 134,306, with 7.8\% women) and biographies of people born after that year ($N_a=$ 456,435, with 16.8\% women).

\subsubsection{Number of Language Editions}

We measure the ratio between men and women as a function of the number of language editions in which they are depicted. 
If the Wikipedia entry indeed functions as a glass ceiling, we expect to see a larger gender gap for ``local heroes'' than for ``superstars,'' because fewer female local heroes would be able to overcome the glass ceiling. The exclusion of less notable women would also imply that, on average, women in Wikipedia should be more notable than their male counterparts. On the contrary, the inclusion of less notable men would decrease the average notability of men in Wikipedia.

\begin{figure}[b]
\centering
 \begin{tabular}{c}
\includegraphics[width=0.95\linewidth]{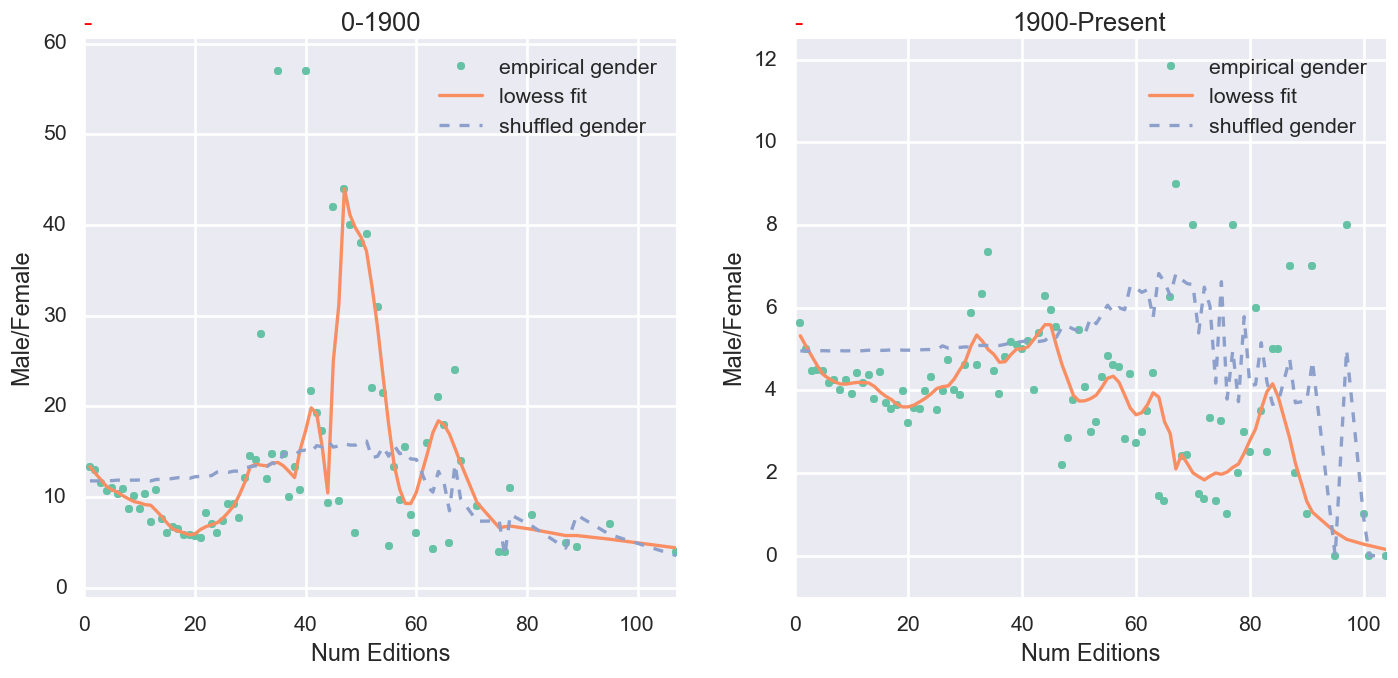}
\end{tabular}
\caption{\textbf{Men-Women Ratio:} 
Ratio of men to women that show up in N language editions before 1900 (Left) and in/after 1900 (Right). 
In and after 1900 the gap between the number of men and women is larger for people with low or medium level of global notability than for the global superstars. The empirical observed gender gap for people with low and medium notability goes beyond what we would expect by randomly reshuffling the gender of people.}
\label{fig:numlang_ratio}
\end{figure}

Figure~\ref{fig:numlang_ratio} shows that since 1900, the gap between men and women is indeed larger for people with low or medium level of global notability than for the ``global superstars.'' 
Especially for local heroes (i.e., people who are only depicted in 1 language edition), the men to women ratio is larger than expect by chance.  In the population of people born after 1900 men are 5.62 times more likely than women to be depicted in Wikipedia if they are only relevant for one specific language community. By random chance (i.e. when we reshuffel the gender) we would expect that men are only 4.94 times more likely than women to be depicted in exactly 1 language edition. That means women are around 14\% less likely to be depicted as local heroes than we would expect by chance.
This finding is important, since almost half of our population belongs to the group of local heroes (45\% of men and 40\% of women are depicted only in 1 language edition).

Also for people born before 1900 we see that the observed men-women ratio for local heroes (13,28) goes beyond what we would expect by chance (11,73). In this case women are around 13\% less likely to be depicted as local heroes than we would expect by chance. Again, a large portion of our population belongs to this group (44\% of men and 39\% of women).
The main difference between our 2 populations is that the gender gap for people born before 1900 does not decrease systematically with increasing notability. 

A possible explanation for the high men-to-women ratio for local heroes is that the entry barrier into Wikipedia is higher for women than for men.
Note that people can also create articles about themselves in Wikipedia; men are on average more self-absorbed than women \cite{Grijalva2015}, and thus may be more likely to create articles about themselves.
Another possible explanation is that more information may be available online about less notable men than about less notable women. Since Wikipedia editors rely on secondary information sources, their decisions also reflect the biases that exist in other media.
 
To further quantify the glass-ceiling effect while controlling for other factors that may potentially explain our results (\eg, profession and age), we use a negative binomial regression model and explore the effect of gender on the number of language editions including a person.
We performed three different regressions: one for people born before 1900 ($N_b$), one for people born since 1900 ($N_a$), and one for the entire dataset ($N$). 
The coefficients that are reported in Table~\ref{table:edition_count_regression} can be  interpreted as follows: if all other factors in the corresponding model were held constant, an increase of one unit in the factor (\eg, from male to female, from Person to Scientist, etc.) would increase the logarithm of the number of editions by the fitted coefficient $\beta$. The Incidence Rate Ratio (IRR) of each factor is obtained by exponentiating its coefficient. 

The regression from the full dataset (last column in Table~\ref{table:edition_count_regression}) reveals that being female makes a biography increase its edition count by an IRR of $1.13$, all other parameters equal. This effect is significant ($p < 0.001$), indicating that women in Wikipedia are slightly more notable than their male counterparts. 
We also observe interesting differences for professions. For example, being a \textit{philosopher} has the strongest positive effect on being of global importance ($IRR = 4.70$, $p < 0.001$), while being a \textit{journalist} has the strongest negative effect on global importance ($IRR = 0.37$, $p < 0.001$). This indicates that people with certain professions are more likely to be recognized globally if they contributed something, while others are more likely to be recognized locally. 
While we do observe interesting differences among professions, further analysis is necessary to investigate whether professional differences in notability are confounded by the average birth decade. For instance, a quarter of the top 100 historical figures are philosophers \cite{whosbigger}, while journalists are more likely to have become famous in recent years.

The model further indicates that the decade when a person was born is negatively associated with notability ($IRR = 0.99$, $p < 0.001$); the more historic a person is, the more notable they are from a global perspective. This is expected: people from older centuries appear on  Wikipedia because their ideas and actions have transcended time (through secondary sources). Conversely, people of recent fame can be notable in terms of availability of secondary sources, but not necessarily because their ideas will remain valuable in time. 
Interestingly, we find that the birth decade factor has a different effect when we look at people pre-1900 and post-1900. 
For people born before 1900, as with the global dataset, being historic is associated with notability ($IRR_b = 0.98$, $p < 0.001$). 
When we consider people born since 1900 we find that Wikipedia developed a ``recency bias''; people in this group are slightly more notable if they were born more recently ($IRR = 1.01$, $p  = 0.008$). 
A possible explanation is that younger people may benefit from the greater availability of digital information about them or generated by them, making them more likely to be recognized by Wikipedia editors.

We also find that being female has small but significant effects on being of global importance in both datasets, although the pre-1900 effect is negative while the post-1900 is positive.
For people in Wikipedia born before 1900, being a female decreases the chances of notability, as one would predict based on the historical exclusion of women \cite{bridenthal1987becoming}.
Conversely, for people in Wikipedia born since 1900, being female increases the chances of notability.
Due to the noted relation between being historic and global notability (see Figure~\ref{fig:numlang_birth}), we cannot claim a glass-ceiling effect for inclusion in Wikipedia of women born prior to 1900.

\subsubsection{Google Search Trends}

Let us next compare the external notability proxy (based on geographic and temporal search interest) of a random sample of men and women in Wikipedia born since 1900. Table~\ref{table:gtrend_regression_ols} shows that women in Wikipedia are indeed slightly more of interest to the world according to the relative search volume statistics of Google. Both coefficients are positive. 
However, only the coefficient for the number of regions with volume above the Google threshold is significant ($IRR = 1.08$).
The mean number of regions with search volume above the Google threshold is $1.93$ for women, $1.51$ for men; the median is zero for both.
The mean number of months during which we observe a global search volume above the Google threshold is 32 for women, 30 for men. The median number of months is one for women and zero for men.

\begin{figure}[b]
\centering
 \begin{tabular}{c}
\includegraphics[width=0.95\linewidth]{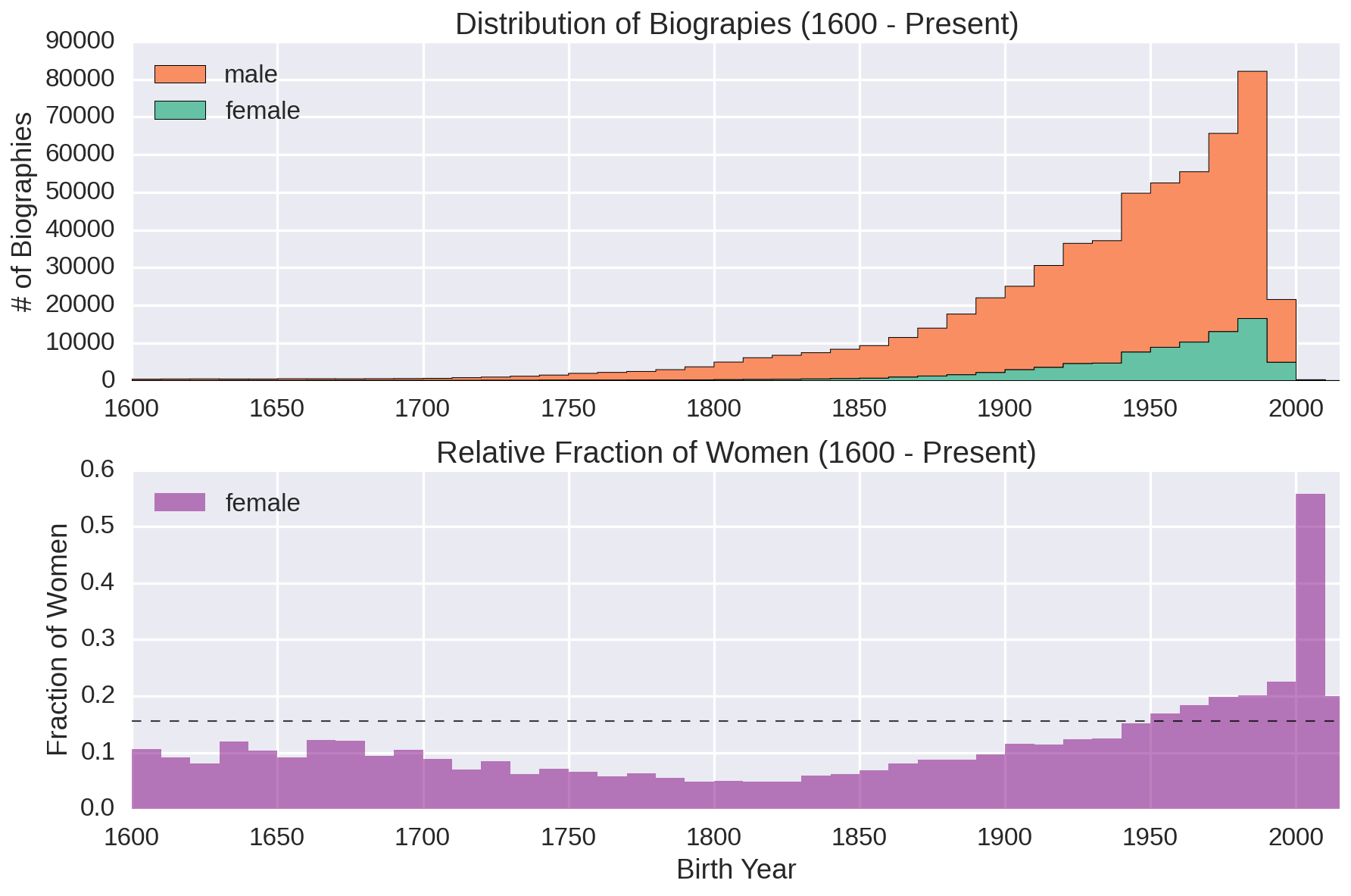}
\end{tabular}
\caption{\textbf{Distribution of Biographies in Time:} The number of men and women in Wikipedia that are born in a certain year. One can see that the number of people that make it to Wikipedia increases with their birth year. Also the fraction of notable women increases.}
\label{fig:time_histogram}
\end{figure}

\begin{figure}[b]
\centering
 \begin{tabular}{c}
\includegraphics[width=0.95\linewidth]{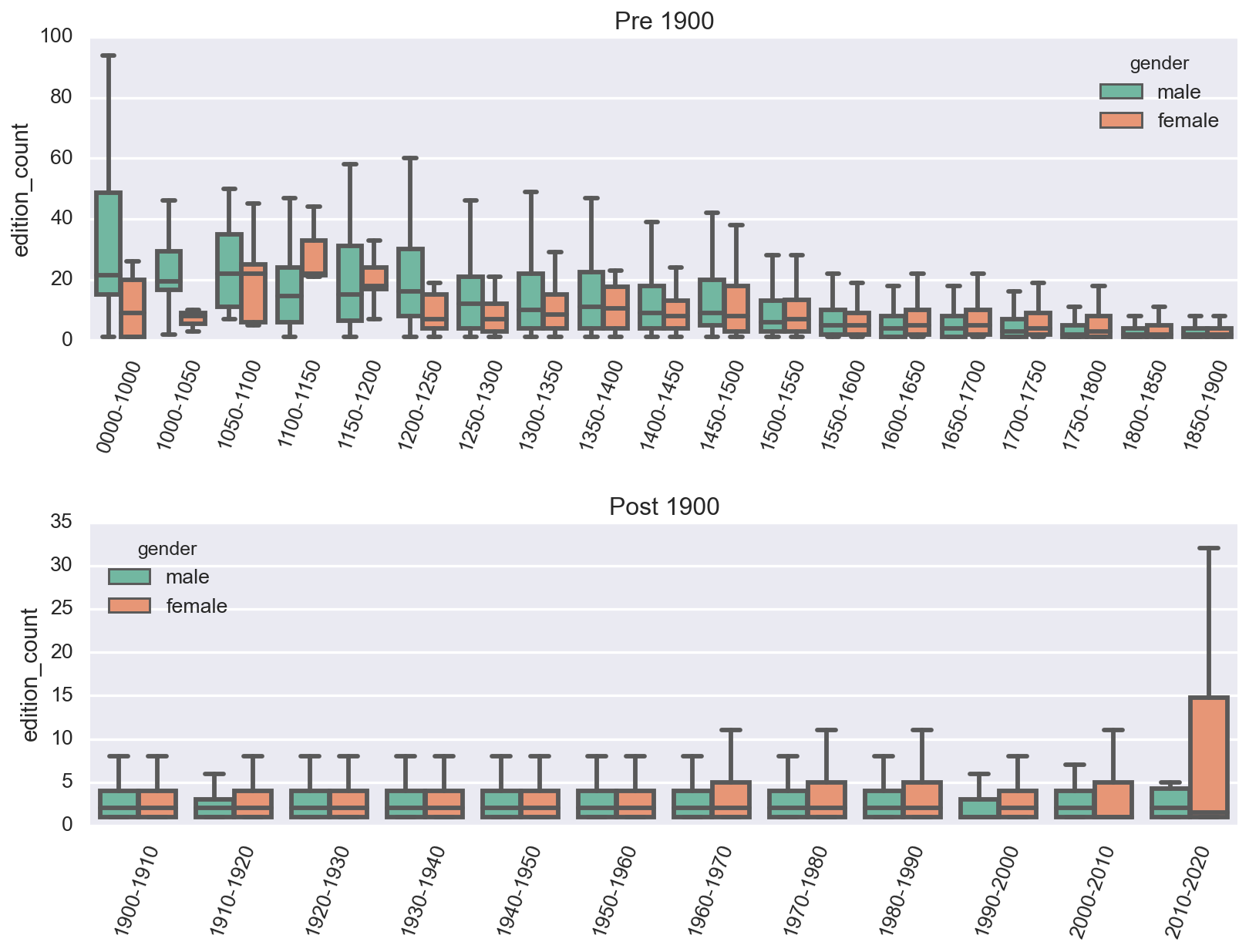}
\end{tabular}
\caption{\textbf{Notability Per Birth Year:} 
The mean number of language editions in which men and women are covered as a function of their birth year. %The shaded region depicts the standard error.
One can see that the global importance is decreasing with birth years which indicates that less historic people are also covered by Wikipedia if they are only of local importance. This can in part be explained by the availability of information about this people, but also by the collective generation process where the editors of each language edition describe their own local heroes. One can see that among the people born after 1600 women are slightly more notable than men, while before 1600 it is the other way around.}
\label{fig:numlang_birth}
\end{figure}

Women included in Wikipedia tend to be born in recent years (see Figure~\ref{fig:time_histogram}) and people born in recent years may have received more attention on Google between 2004 and 2015. 
Controlling for year of birth and profession was not possible due to the technical challenges of collecting large amounts of Google trend data. Focusing on sub-samples of people who are born in the same year and share the same profession may allow to address these confounding factors.

\subsection{Topical and Linguistic Asymmetries}

Language is one of the primary media through which stereotypes are conveyed. We next explore differences in the words and word sequences that are frequently used when writing about men or women to uncover topical and linguistic biases.

\subsubsection{Topical Bias}

Following the notability analysis, we must consider time as a confounding factor. We therefore consider two groups of biographies: those with birth date prior to 1900, and those with birth date from 1900 onwards.
We estimated the PMI of each word and bi-gram in our vocabulary for each gender. Since the PMI give more weight to words with very small frequencies, we considered only n-grams that appear in at least 1\% of men's or women's biography overviews. Our findings for each dataset are summarized as follows:

\begin{itemize}
 \item Pre-1900: the three words most strongly associated with females are \textit{her husband}, \textit{women's}, and \textit{actress}. The three most strongly associated with males are \textit{served}, \textit{elected}, and \textit{politician}.
 \item 1900--onwards: the three words most strongly associated with females are \textit{actress}, \textit{women's}, and \textit{female}. The three most strongly associated with males are \textit{played}, \textit{league}, and \textit{football}.
\end{itemize}

Figure~\ref{fig:wordclouds} shows the n-grams that are strongly associated with each gender. 
The bi-grams that are strongly associated with women born before 1900 relate frequently to categories such as gender, family, and relationships. Words associated with men mainly relate to other categories, such as politics and sports.  
Table~\ref{table:lexical_categories} shows the proportion of the top 200 n-grams that fall into each category, for both genders in both periods. 
The categories \textit{gender}, \textit{relationship}, and {family} are more prominent for women than men. 
However, the distributions of those categories are different in the two periods under consideration. The distribution is significantly different across genders only pre-1900, according to a chi-square test ($\chi^2 = 14.33$, $p < 0.01$). In prior work we have shown that the differences are significant if time is not considered~\cite{graells2015first} and that similar results hold for five other language editions~\cite{wagner2015s}.

\subsubsection{Linguistic Bias}

Table~\ref{table:lib-ratios} shows the ratios of abstract terms among positive and negative terms when aggregating all the text in the summaries of the biographies of men and women separately. One-tailed chi-square tests suggest that linguistic biases appear along the predicted directions: more abstract terms are used for positive aspects of men's biographies and for negative aspects of women's biographies.  Effect sizes, measured by Cohen's $w$, are very small, in line with the typically small effects in other studies in psycholinguistics.
When measuring relative changes, we find that adjectives are almost 9\% more likely to be used to describe positive aspects of men's biographies, while 1.62\% more likely to describe negative aspects in women's biographies.

We apply linear regression in two models, one with $r_{+}$ as dependent
variable and another one with $r_{-}$, expressed as a linear combination of gender,
class, and century of birth. We focus on all biographies with valid birth dates and at least 250 words in their summary. Our results indicate that women's biographies tend to have fewer abstract terms for positive aspects and more abstract terms for negative aspects, as predicted by the LIB (see Table~\ref{table:lib-regression}). This effect is robust to the inclusion of control variables like profession and century of birth. We repeated the analysis using a logit transformation of $r_{-}$ and $r_{+}$, as well as with beta regression, finding the same results.

\subsection{Structural Inequalities}

Structured information in Wikipedia serves many purposes, from providing input data to search engines, to feeding knowledge databases. Thus, inequalities in structure have an influence that goes beyond Wikipedia, regardless of being a reflection of society or history, or being inherent to Wikipedia contributors.

\subsubsection{Meta-data}

In total, the DBpedia dataset contains 340 attributes extracted from infobox templates. Of those attributes, 33 display statistically significant differences. Only 14 of them are present in at least 1\% of the male or female biographies.
These attributes are shown in Table~\ref{table:biography-meta-data}.
As in previous sections, we have estimated the significance of their differences for people born before and since 1900. An analysis of the entire dataset without considering time is presented in our previous work \cite{graells2015first}.

Due to the number of available attributes, the portion of biographies that contains each of them is small. Thus, instead of considering $p$-value correction, we discuss the statistically significant gender differences manifested in the meta-data to qualitatively assess whether they have significance in our context:

\begin{itemize}
\item Attributes \textit{activeYearsEndDate}, \textit{activeYearsStartYear}, \textit{careerStation}, \textit{numberOfMatches}, \textit{position}, \textit{team}, and \textit{years} are more frequently used to describe men.
All of these attributes are related to sports, therefore the differences can be explained by the prominence of men in sports-related DBpedia classes (\eg, \textit{Athlete}, \textit{SportsManager} and \textit{Coach} \cite{graells2015first}).
Differences in \textit{activeYearsStartYear} are only significant at the entire dataset level, and differences in \textit{activeYearsEndDate} are only significant before the 20th century. The other attributes are mostly significantly different in recent times.

\item Attributes \textit{deathDate} and \textit{deathYear} are more frequently used for men born before 1900. A possible explanation is that the life of women was less well documented  than the life of men in the past, and therefore it is more likely that the death date or birth date is unknown for women.

\item Attribute \textit{birthName} is more frequently used for women in recent times. 
Its value refer mostly to the original name of artists, and women have considerable presence in this class \cite{graells2015first}. 
A likely explanation is that married women change their surnames to those of their husbands in some cultures. 

\item Attributes \textit{occupation} and \textit{title} are more frequently used to describe women in recent times, and seem to serve the same purpose but through different mechanisms. On one hand, \textit{title} is a text description of a person's occupation (the most common values found are \textit{Actor} and \textit{Actress}). On the other hand, \textit{occupation} is a DBpedia resource URI (\eg, \url{http://dbpedia.org/resource/Actress}). 
These attributes are present in the infoboxes of art-related biographies. Conversely, the infoboxes of sport-related biographies do not contain these attributes because their templates are different and contain other attributes (like the aforementioned \textit{careerStation} and \textit{position}). Thus the meta-data of athletes, who are mostly men, do not contain such attributes.

\item The \textit{homepage} attribute is more frequently used for women in recent times. Our manual inspection showed that biographies from the \textit{Artist} class tend to have homepages, which explains why the attribute is used more frequently for women.

\item The \textit{spouse} attribute is more frequently used for women in recent times. 
This attribute indicates whether the portrayed person was married or not, and with whom. 
In some cases, it contains the resource URI of the spouse, while in other cases, it contains the name (\ie, when the spouse does not have a Wikipedia article), or the resource URI of the article of \textit{``divorced status.''}
This difference is consistent with our results about topical gender difference, where terms related to relationships show a stronger association with women than men.
\end{itemize}

All differences found have large effect sizes (Cohen's $w > 0.5$).

\subsubsection{Network Structure}

We constructed the empirical network from the inter-article links among 893,380 biographical articles in the English Wikipedia.
After removing 192,674 singleton nodes (of which 15.3\% were female), the resulting graph had $n=$ 700,706 nodes (of which 15.6\% were female) and 4,153,978 edges. 
All baseline graphs have the same number of nodes $n$ and approximately the same mean degree $k \approx 4$ as the empirical network.
The small world baseline has a parameter $\beta = 0.34$ representing the probability of rewiring each edge. Its value was set using the Brent root finding method in such a way as to recover the clustering coefficient of the original network.

\begin{figure}[tb]
\centering
\includegraphics[width=\linewidth]{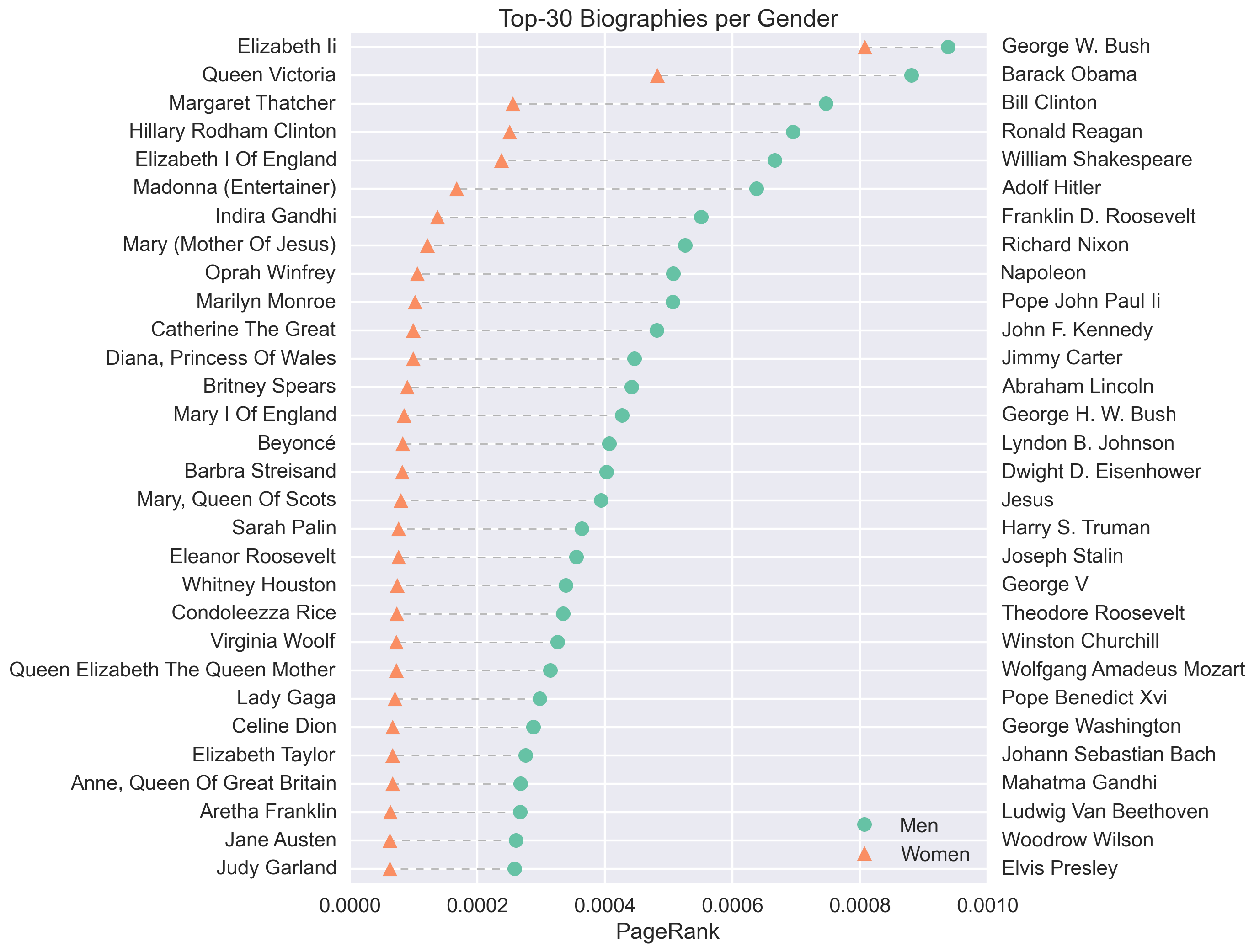}
\caption{\textbf{Top-30 biographies sorted by Page Rank}. One can see that women are slightly less central than men and also centrality of women decreases faster with decreasing rank than the centrality of men.}
\label{fig:gender_pagerank}
\end{figure}

Figure~\ref{fig:gender_pagerank} shows the top 30 men and women according to their PageRank. 
The top-ranked women are slightly less central than men, and the centrality of women decreases faster than that of men with decreasing rank.
The top-ranked biographies are similar to those found in previous work~\cite{aragon2012biographical,whosbigger}.

In addition to the full hyperlink network, we created two sub-networks: one only contains people born before 1900 and the other only contains people born since 1900.
For each empirical network, we created several null models and compared the proportion of links within and across genders using a chi-square test.
Table~\ref{table:biography-network-properties} indicates that in both
empirically observed Wikipedia graphs, women biographies have more links to other women articles than one would expect by chance. 
A possible explanation for this asymmetry stems from the reported interests of female editors, who frequently edit biographies about women in Wikipedia \cite{wikisurvey}.

\begin{figure}[tb]
\centering
\includegraphics[width=\linewidth]{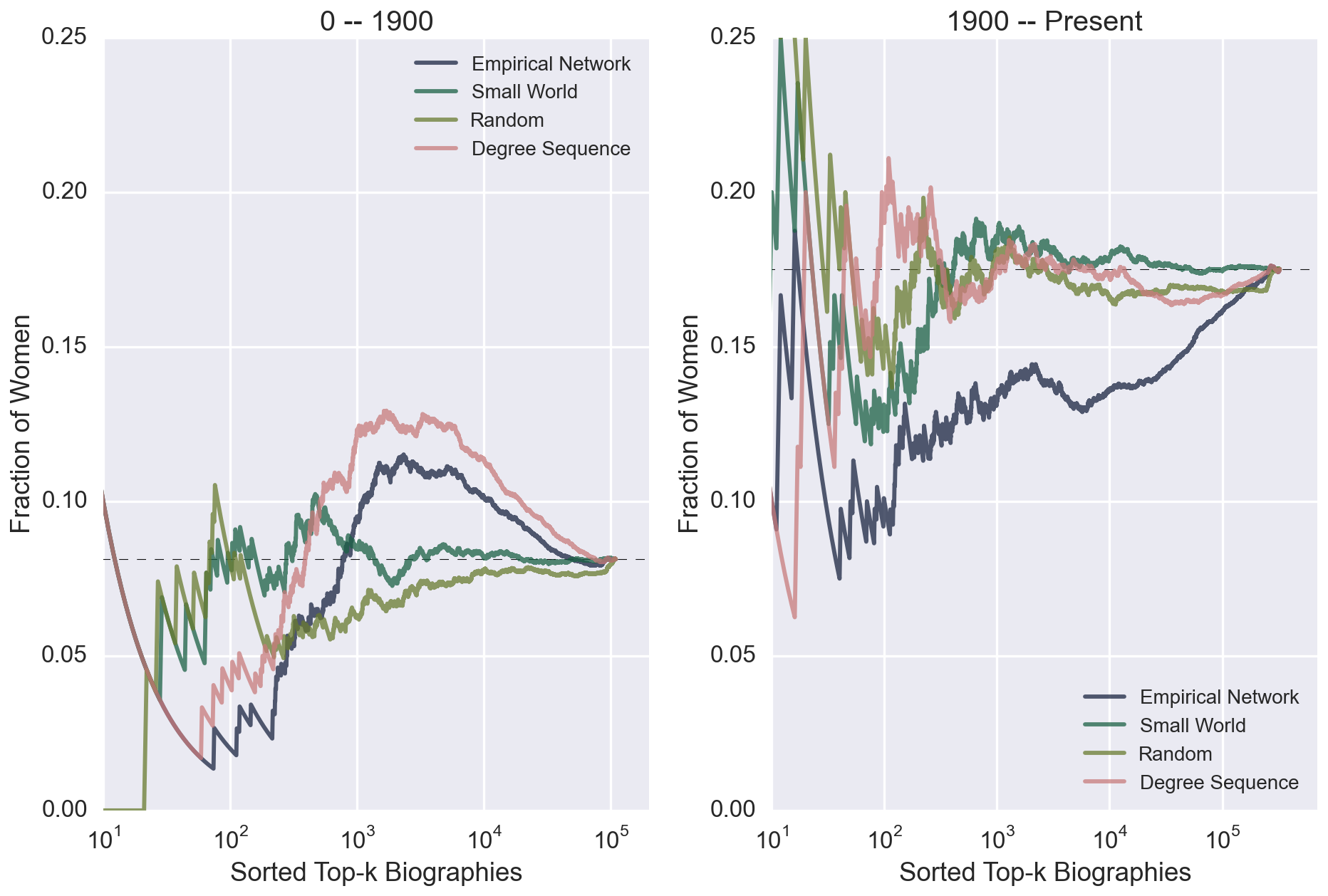}
\caption{
\textbf{Women fraction in top $k$ biographies sorted by PageRank.} 
One can see that the relative fraction of women in the top $k$ biographies in the three baseline networks converges faster to the expected fraction than in the empirical observed network (OBS) where nodes are ranked by Page Rank and Indegree. This indicates, that the empirical observed topology of the hyperlink network puts women (especially women born in 1900 or afterward) in an disadvantage when it comes to ranking algorithms. 
}
\label{fig:ccdf-pagerank}
\end{figure}

\begin{figure}[b]
\centering
 \begin{tabular}{c}
        \includegraphics[width=0.9\linewidth]{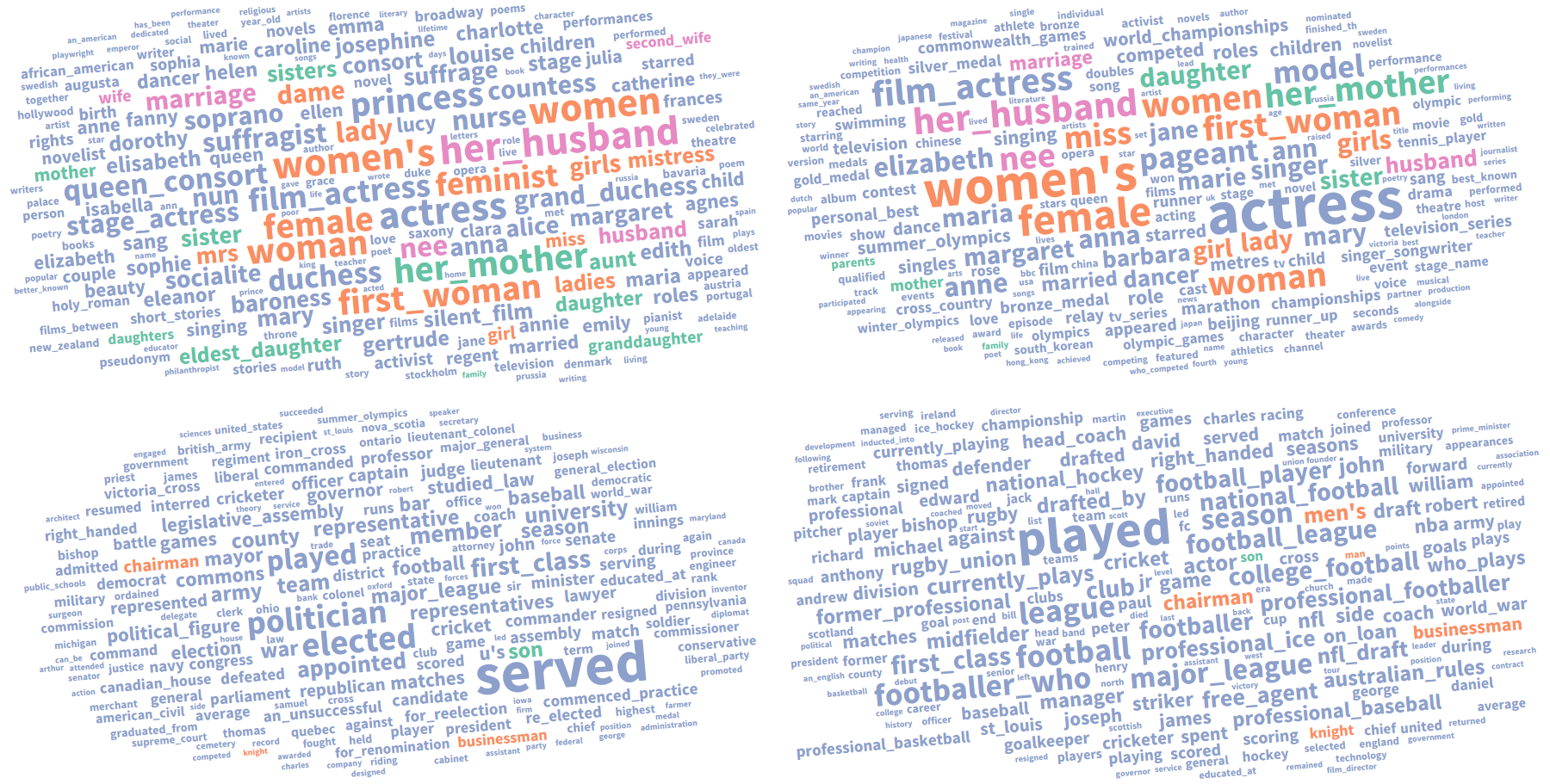}
        \end{tabular}
        \caption{\textbf{Topical Bias:} Word clouds for articles about women (top) and men (bottom), separated by time: biographies with birth date before 1900 are on the left, and after (including) 1900, on the right. Font size is proportional to PMI with each gender. The color depicts the four categories (\textit{gender} --orange--, \textit{family} --green--, \textit{relationship} --violet--, and \textit{other} --blue--). One can see that beside professional and topical areas, words that fall into the category gender, relationship and family are more dominant in articles about women born before 1900. Gender-specific differences are much less pronounced in articles about people born in or after 1900.}
\label{fig:wordclouds}
\end{figure}

The effect of structural differences on visibility can be analyzed in terms of how many women are ranked among the top biographies by centrality scores. Figure~\ref{fig:ccdf-pagerank} displays the  fraction of women in subsets of top-ranked biographies.
For people born before 1900, the fraction of women in the top $k$ biographies is below the expected ratio of 7.8\% up to $k \approx 10^3$, and above when lower-ranked biographies are considered. For people born since 1900, the fraction or women is below the expected ratio of 16.8\% for the entire range of $k$. 
This indicates that the empirically observed structure of the Wikipedia hyperlink network puts women at a disadvantage when it comes to ranking algorithms, especially for women born since 1900. 
For people born before 1900, as $k$ increases, the relative fractions of women among the top $k$ biographies in the baseline networks converge to the expected ratios faster than in the empirical networks. 
This implies an asymmetry that cannot simply be explained by heterogeneities in the structure of the networks, since our baseline graphs preserve several characteristics of the empirical network, including the broad distribution of node degrees. Therefore one must conclude that there exists a bias in the generation of links by Wikipedia editors, favoring articles about men.

\section{Discussion}

In previous work we found that notable women and men from three different reference lists have equal probability of being represented in Wikipedia \cite{wagner2015s}.
While this result is encouraging, external reference lists may also be biased. 
For example, if women that show up in these reference lists are more notable than their male counterparts, then equality in coverage does not imply the absence of gender bias. 
However, assessing the notability of people is a difficult task. In this work we propose to use Wikipedia edits in different language editions and search engines like Google to estimate the public interest in a person at different times and in different regions. Wikipedia view statistics could be used to extend this internal proxy measure of notability in the future.

Our analysis of the global notability of men and women in Wikipedia reveals that women are slightly more notable than men, even if we control for confounding factors such as professions (\eg, philosophers have high global notability and most of them are men) and year of birth (historic people are more notable and until recently our history was dominated by men). 
Further, the ratio of women is smaller for low levels of notability than for high levels.
These findings suggest the existence of a subtle glass-ceiling effect that makes it more difficult for women to be included in Wikipedia than for men.
At least three plausible explanations exist that describe why the glass-ceiling effect may be present in Wikipedia:  (1) the narrow diversity of editors may foster the glass-ceiling effect since it is well known that individuals generally favor people from their in-group over people from their out-group \cite{Brown1995,Tajfel1971}; (2) men are potentially more likely to create an article about themselves since previous research suggests that men are on average more self-absorbed than women \cite{Grijalva2015}; (3) the external materials on which Wikipedia editors rely may introduce this bias, since the life of women or certain ethnic minorities may be less well documented and less visible on the Web.
We leave the question of identifying what causes this effect for future research.

One way to mitigate the glass-ceiling effect is by relaxing notability guidelines for women, in order to include women who are locally notable, and for whom secondary sources might be hard to find.
We acknowledge that this is not easy, because relaxing notability guidelines can open the door for original research, which is not allowed in Wikipedia. 
However, a well-defined affirmative strategy would allow for the proportion of women in Wikipedia to grow and make women easier to find, alleviating 
several asymmetries found. 

The topical and  linguistic asymmetries that we found highlight that editors need to pay attention to the ways women are portrayed in Wikipedia. 
Critics may rightly say that by relying on secondary sources, Wikipedia just reflects the biases found in them.
However, editors are expected to write in their own words ``while substantially retaining the meaning of the source material''\footnote{\url{https://en.wikipedia.org/wiki/Wikipedia:No_original_research}} and thus, the differences found in terms of language are caused explicitly by them.
Efforts to mitigate linguistic bias could include a revision of the neutral point of view (NPOV) guidelines\footnote{\url{https://en.wikipedia.org/wiki/Wikipedia:Neutral_point_of_view}} to explicitly address gender bias. A simple example would be the Finkbeiner test: does the article mention the person's gender? Is it needed?

Even though the structural inequalities that we found suggest that editors (especially those who edit articles about women) do a great job in interlinking articles about women, the visibility of women is still lower than expected when link-based ranking algorithms such as PageRank are applied. Since the majority of biographies are about men and men tend to link more to men than to women (see Figure 6 in \cite{eom2014} for preliminary comparison of ranking algorithms),
future research should focus on developing search and ranking algorithms that account for potential discrimination of minority groups due to homophily, \ie, the tendency of nodes to link to similar nodes.

Wikipedia should provide tools to help editors, for instance, by considering already existing manuals of gender-neutral language \cite{apa_manual}, or by indicating missing links between articles. For example, if an article about a woman links to the article about her husband, the husband should also link back. 
Internal Wikipedia discussions that started after we published our preliminary studies on gender inequalities in the content of Wikipedia \cite{wagner2015s,graells2015first} suggest such actions\footnote{\url{https://en.wikipedia.org/wiki/Wikipedia:Writing_about_women}}. However they are not yet internal policies.

\section{Related Work}

\textbf{Gender Inequalities in Traditional Media:} Feminists often claim that news is not just mostly about men, but overwhelmingly seen through the eyes of men.
Analysis of longitudinal data from the Global Media Monitoring Project (GMMP) spanning over 15 years indicates that the role of women as producers and subjects of news has seen a steady improvement, but the relative visibility of women compared to men has been stuck at 1:3 \cite{Ross2011}.
Gender inequalities are also manifested in films used for education purposes, as revealed by the application of the Bechdel test to teaching content \cite{Scheiner-Fisher2012}.

\textbf{Gender Inequalities in Wikipedia:}
Our work is not the first to recognize the importance of understanding gender biases in Wikipedia \cite{reagle2011gender,eom2014,callahan2011cultural,aragon2012biographical,wagner2015s,graells2015first}.

Reagle and Lauren \cite{reagle2011gender} compare the  coverage and article length of thousands of biographical subjects from six reference sources (\eg, \textit{The Atlantic}'s 100 most influential figures in American history, \textit{TIME Magazine}'s list of
2008's most influential people) in the English-language Wikipedia and the online Encyclopedia Britannica. 
The authors do not find gender-specific differences in the coverage and article length in Wikipedia, but Wikipedia’s missing articles are disproportionately female relative to those of Britannica. 
Wagner \textit{et al.}~\cite{wagner2015s} also analyzed the coverage of notable people in Wikipedia based on three external reference lists (Pantheon \cite{Pantheon}, Freebase \cite{Schich2014} and Human Accomplishment \cite{Murray2003}) and found no significant difference  in the proportional coverage of men and women in six different language edition of Wikipedia.

Bamman and Smith \cite{bamman2014unsupervised} present a method to learn biographical structures from text and observe that in the English Wikipedia, the biographies of women disproportionately focus on marriage and divorce compared to those of men, in line with our findings on the lexical dimension.
Similar results are found by Graells-Garrido \textit{et al.}~\cite{graells2015first} where the most important n-grams and LIWC categories of men and women are compared. 
Similar topical biases are found in six different language editions (German, English, French, Italian, Spanish and Russian) \cite{wagner2015s}.

Recent research shows that most important historical figures across Wikipedia language
editions are born in western countries after the 17th century, and are male \cite{eom2014}. 
The authors use different link-based ranking algorithms and focus on the top 100 figures in each language edition. 
Their results show that very few women are among the top 100 figures --- 5.2 on average across language editions. Since the authors do not use external reference lists, it remains unclear how many women we would expect to see among the top 100 figures.

In terms of network structure, we built a biography network \cite{aragon2012biographical} in which we estimated PageRank, a measure of node centrality based on network connectivity~\cite{brin1998anatomy,Fortunato08internetmath}. 
In similar contexts, PageRank has been used to provide an approximation of historical importance \cite{aragon2012biographical,whosbigger} and to study the bias leading to the gender gap~\cite{whosbigger}.

Previous research has also explored gender inequalities in the editor community of Wikipedia and potential reasons   \cite{lam2011wp,Collier2012,hill2013wikipedia}.
The importance of this issue has been acknowledged among Wikipedians,  for example through the initiation of the ``Countering Systemic Bias'' WikiProject\footnote{\url{http://en.wikipedia.org/wiki/Wikipedia:WikiProject_Countering_systemic_bias}} in 2004.

Though previous research identified gender bias on a topical and structural level in Wikipedia, the present work goes beyond previous efforts by (i)~providing an in-depth analysis of the content and structure of the English Wikipedia, (ii)~analyzing external and internal signals of global notability of men and women that are depicted in Wikipedia, and (iii) exploring to what extent linguistic biases manifest in the content of Wikipedia.

\section{Conclusions}

In this paper we studied various aspects of gender bias in the content of Wikipedia biographies. This is an important issue since the usage of Wikipedia is growing, and with that, its importance as a central knowledge repository that is used around the globe, including for educational purposes.

Our empirical results uncover significant gender differences at various levels that cannot only be attributed to the fact that Wikipedia is mirroring the off-line world and its biases.
For instance, the lexical and linguistic differences must be attributed to Wikipedia editors, since they are expected to use their own words.
We believe that the differences in the notability of men and women that are present in Wikipedia can in part be explained by how the life of men and women is documented in our society \cite{Thomas1992}.
Since Wikipedia editors do rely on this biased information for informing their decisions (\eg, who is notable enough to be depicted in Wikipedia? What are the most important facts about this person?), it is not surprising that the content they produce reflects these pre-existing biases.
However, it is also well known from social psychology that human-beings generally favor people in their in-group over people in their out-group \cite{Brown1995,Tajfel1971} and our results show that Wikipedia editors reveal a linguistic in-group/out-group bias \cite{Maass1989}. 

The extent to which this bias also impacts the selection (or article creation) process of notable people remains however unclear.  Interestingly, we find that women that are depicted in Wikipedia tend to be more notable than men from a global perspective, which can be seen as an indication of gender-specific entry barriers.

Our empirical results are limited to the English Wikipedia, which is biased towards western cultures \cite{hecht2009measuring}.
However, in previous work \cite{wagner2015s} we found that similar structural, topical and coverage biases exist across six different language editions. 
We leave a more detailed exploration of gender bias across all language editions for future work.
Our methods can be applied in other contexts given an ad-hoc manual coding of associated keywords to each gender.

In summary, the contributions of this work are twofold: (i) we presented a computational method for assessing gender bias in Wikipedia \emph{along multiple dimensions} and (ii) we applied this method to the English Wikipedia and shared empirical insights on observed gender inequalities. 
The methods presented in this work can be used to assess, monitor and evaluate these issues in Wikipedia on an ongoing basis. 
We translate our findings into some potential actions for the Wikipedia editor community to reduce gender biases in the future. We hope our work will contribute to increased awareness about gender biases online, and about the different ways these biases can manifest themselves.
We propose that Wikipedia may wish to consider revising its guidelines, both to account for the low visibility of women and to encourage a less biased use of language.

\spara{Competing interests}
The authors declare that they have no competing interests.

\spara{Author's contributions}
All authors contributed to the research design and writing of the paper.
Claudia Wagner was mainly responsible for the internal and external notability study and the topical analysis.
Eduardo Graells-Garrido was collecting and preparing the data. Further he was working on the internal notability study, the network and topic analyses.
David Garcia focused on the linguistic bias exploration. 
Filippo Menczer was mainly responsible for the network analysis.

\spara{Acknowledgments}
We thank Mounia Lalmas, Markus Strohmaier, and Mohsen Jadidi for their valuable input to this research.

\begin{table}[tb]
\centering
\scriptsize
\caption{\textbf{The largest 20 language editions of Wikipedia}:
The number of biographies, proportion of biographies about women, and overlap with biographies in the English edition are depicted. The fraction of women is on average around 17\% and the average overlap with English is 97\%.}
\begin{tabular}{lccr}
\toprule
Language & Fraction of Women & Overlap with English Edition & Biographies \\                       
\midrule
English (en)  &    0.155   &     --    &    893,380  \\
Italian (it)  &    0.151   &     0.986    &    134,122 \\
Deutsch (de)  &    0.132   &     0.995    &    102,233 \\
French (fr) &     0.136    &     0.966      &  93,400 \\
Polish (pl)  &    0.158    &     0.986     &   69,531\\
Spanish (es) &     0.182   &     0.980    &    66,067\\
Russian (ru) &     0.158   &     0.988    &    64,233\\
Portuguese (pt)  &    0.185 &        0.989 &       44,793\\ 
Dutch (nl)  &    0.194     &   0.993      &  38,659 \\
Japanese (ja) &     0.184  &      0.991   &     31,033 \\
Hungarian (hu)  &    0.179 &       0.999  &      18,074 \\
Bulgarian (bg)  &    0.149    &    1.000      &  16,850 \\
Korean (ko)   &   0.226    &    0.994  &      15,921\\ 
Turkish (tr)  &    0.175   &     0.982 &       14,399 \\
Indonesian (id)  &    0.151   &     0.987      &  12,401 \\
Arabic (ar) & 0.199 & 0.787 & 12,030 \\
Czech (cs)  &    0.156   &     1.000      &  10,765 \\
Catalan (ca)  &    0.183   &     0.995 &       7,721\\ 
Greek (el) & 0.145 & 0.806 & 6,748 \\
Basque (eu)   &   0.179    &    0.987  &      3,449  \\
\bottomrule
\end{tabular}

\label{table:dataset}
\end{table}

\begin{table}[tb]
\centering
\scriptsize
\caption{\textbf{Interest via Number of Language Editions}: Results of three negative binomial regression models that use the number of language editions in which an article about a person shows up as dependent variable and gender as independent variable, while controlling for profession and birth century. In the full dataset and the subset of people born after 1900, women are slightly more notable than men since the coefficient is significantly positive even when controlling for other variables such as professions and age. $^{***}: p < 0.001$}
\begin{tabular}{l|ccc|ccc|ccc}
\toprule
    & \multicolumn{3}{c|}{0 -- 1899} & \multicolumn{3}{c|}{1900 -- Present} & \multicolumn{3}{c}{0 -- Present} \\
    &      $\beta$ & std. err. & p &      $\beta$ & std. err. & p &      $\beta$ & std. err. & p  \\
\midrule

C(class)[T.Ambassador]         &  0.083 & 0.148 & 0.574 & -0.537 & 0.076 & $^{***}$ & -0.412 & 0.068 & $^{***}$ \\
C(class)[T.Architect]          &  0.355 & 0.041 & $^{***}$ &  0.574 & 0.047 & $^{***}$ &  0.421 & 0.031 & $^{***}$ \\
C(class)[T.Artist]             &  0.853 & 0.012 & $^{***}$ &  0.420 & 0.005 & $^{***}$ &  0.508 & 0.005 & $^{***}$ \\
C(class)[T.Astronaut]          &    -- &   -- &   -- &  1.403 & 0.038 & $^{***}$ &  1.428 & 0.038 & $^{***}$ \\
C(class)[T.Athlete]            & -0.344 & 0.011 & $^{***}$ &  0.042 & 0.004 & $^{***}$ &  0.084 & 0.003 & $^{***}$ \\
C(class)[T.BeautyQueen]        &    -- &   -- &   -- & -0.290 & 0.035 & $^{***}$ & -0.206 & 0.035 & $^{***}$ \\
C(class)[T.BusinessPerson]     & -1.066 & 0.254 & $^{***}$ & -0.929 & 0.173 & $^{***}$ & -0.983 & 0.143 & $^{***}$ \\
C(class)[T.Chef]               &  0.272 & 0.571 & 0.633 & -0.268 & 0.070 & $^{***}$ & -0.217 & 0.070 & 0.002 \\
C(class)[T.Cleric]             &  0.545 & 0.022 & $^{***}$ &  0.417 & 0.020 & $^{***}$ &  0.477 & 0.015 & $^{***}$ \\
C(class)[T.Coach]              & -0.932 & 0.042 & $^{***}$ & -0.938 & 0.023 & $^{***}$ & -0.941 & 0.020 & $^{***}$ \\
C(class)[T.Criminal]           &  0.468 & 0.073 & $^{***}$ &  0.197 & 0.030 & $^{***}$ &  0.244 & 0.028 & $^{***}$ \\
C(class)[T.Economist]          &  1.504 & 0.099 & $^{***}$ &  0.941 & 0.045 & $^{***}$ &  1.043 & 0.041 & $^{***}$ \\
C(class)[T.Engineer]           &  0.411 & 0.054 & $^{***}$ &  0.002 & 0.079 & 0.979 &  0.243 & 0.044 & $^{***}$ \\
C(class)[T.FictionalCharacter] &    -- &   -- &   -- & -1.021 & 0.418 & 0.015 & -0.969 & 0.419 & 0.021 \\
C(class)[T.Historian]          & -0.579 & 0.172 & 0.001 & -0.756 & 0.117 & $^{***}$ & -0.730 & 0.097 & $^{***}$ \\
C(class)[T.HorseTrainer]       & -0.983 & 0.563 & 0.081 & -0.999 & 0.107 & $^{***}$ & -0.987 & 0.106 & $^{***}$ \\
C(class)[T.Journalist]         & -0.899 & 0.176 & $^{***}$ & -1.032 & 0.078 & $^{***}$ & -1.005 & 0.072 & $^{***}$ \\
C(class)[T.Judge]              & -0.580 & 0.055 & $^{***}$ & -0.700 & 0.040 & $^{***}$ & -0.677 & 0.033 & $^{***}$ \\
C(class)[T.MilitaryPerson]     & -0.014 & 0.011 & 0.195 & -0.287 & 0.013 & $^{***}$ & -0.166 & 0.008 & $^{***}$ \\
C(class)[T.Model]              & -0.146 & 0.704 & 0.836 &  0.249 & 0.030 & $^{***}$ &  0.332 & 0.030 & $^{***}$ \\
C(class)[T.Monarch]            &  1.024 & 0.064 & $^{***}$ &  1.313 & 0.119 & $^{***}$ &  1.227 & 0.056 & $^{***}$ \\
C(class)[T.Noble]              &  0.096 & 0.029 & 0.001 &  0.009 & 0.135 & 0.944 &  0.175 & 0.028 & $^{***}$ \\
C(class)[T.OfficeHolder]       &  0.340 & 0.011 & $^{***}$ &  0.300 & 0.007 & $^{***}$ &  0.308 & 0.006 & $^{***}$ \\
C(class)[T.Philosopher]        &  1.992 & 0.050 & $^{***}$ &  1.180 & 0.040 & $^{***}$ &  1.547 & 0.031 & $^{***}$ \\
C(class)[T.PlayboyPlaymate]    &    -- &   -- &   -- & -0.068 & 0.078 & 0.381 & -0.014 & 0.078 & 0.854 \\
C(class)[T.Politician]         &  0.067 & 0.011 & $^{***}$ &  0.098 & 0.009 & $^{***}$ &  0.068 & 0.007 & $^{***}$ \\
C(class)[T.Presenter]          &  0.121 & 0.458 & 0.792 & -0.758 & 0.068 & $^{***}$ & -0.701 & 0.068 & $^{***}$ \\
C(class)[T.Religious]          &  0.295 & 0.115 & 0.010 &  0.112 & 0.076 & 0.145 &  0.172 & 0.064 & 0.007 \\
C(class)[T.Royalty]            &  1.175 & 0.017 & $^{***}$ &  1.077 & 0.029 & $^{***}$ &  1.155 & 0.015 & $^{***}$ \\
C(class)[T.Scientist]          &  1.191 & 0.014 & $^{***}$ &  0.631 & 0.012 & $^{***}$ &  0.854 & 0.009 & $^{***}$ \\
C(class)[T.SportsManager]      &  0.306 & 0.053 & $^{***}$ &  0.464 & 0.010 & $^{***}$ &  0.493 & 0.010 & $^{***}$ \\
C(gender)[T.female]            & -0.044 & 0.011 & $^{***}$ &  0.116 & 0.004 & $^{***}$ &  0.119 & 0.004 & $^{***}$ \\
birth\_decade     & -0.017 & 0.000 & $^{***}$ &  0.010 & 0.001 & $^{***}$ & -0.010 & 0.000 & $^{***}$ \\

Intercept                    &  4.269 & 0.060 & $^{***}$ & -0.684 & 0.131 & $^{***}$ &  3.022 & 0.038 & $^{***}$ \\
%alpha          &  0.750 & 0.004 & $^{***}$ &  0.754 & 0.002 & $^{***}$ &  0.762 & 0.002 & $^{***}$ \\
\midrule
AIC & \multicolumn{3}{c|}{660,646.944}  & \multicolumn{3}{c|}{2,206,624.237} & \multicolumn{3}{c}{2,873,689.603} \\
%BIC & \multicolumn{3}{c|}{660,950.988}  & \multicolumn{3}{c|}{2,207,010.329} & \multicolumn{3}{c}{2,874,084.722} \\
%Log-Likelihood & \multicolumn{3}{c|}{-330,292.472}  & \multicolumn{3}{c|}{-1,103,277.119} & \multicolumn{3}{c}{-1,436,809.801}\\
%Deviance & \multicolumn{3}{c|}{2,246,465.353}  & \multicolumn{3}{c|}{8,154,463.762} & \multicolumn{3}{c}{10,724,068.586}\\
%LL Ratio & \multicolumn{3}{c|}{30,749.738}  & \multicolumn{3}{c|}{21,759.407} & \multicolumn{3}{c}{49,404.810}\\
Num. obs. & \multicolumn{3}{c|}{134,306.000}  & \multicolumn{3}{c|}{456,435.000} & \multicolumn{3}{c}{590,741.000} \\
\bottomrule
\end{tabular}

\label{table:edition_count_regression}
\end{table}

\begin{table}[tb]
\centering
\scriptsize
\caption{\textbf{Interest via Google Trend Data:} Linear regression results where the number of regions or number of months with a search volume above the threshold were used as independent variable and gender is used as dependent variable. We use a random sample of 5245 people born after 1900 and before 2000 to fit the model. One can see that women inside Wikipedia are on average of interest to people from more different geographic regions than men. The difference in the number of months in which men and women expose a global search volume above google's threshold is not significant.  $^{***}: p < 0.001, ^{*}: p < 0.05$}
\label{table:gtrend_regression_ols}
\begin{tabular}{l|ccc|ccc}
\toprule
    & \multicolumn{3}{c|}{Num Regions} & \multicolumn{3}{c|}{Num Months} \\
    & $\beta$ & std. err. & p &  $\beta$ & std. err. & p  \\
\midrule
Intercept   &   1.5090  &        0.083 &  $^{***}$  &   3.3880  &        0.022   &   $^{***}$   \\
C(gender)[T.female]  & 0.4179  &        0.209   &  $^{*}$  &  0.0978  &        0.056 & $0.081 $ \\
\hline
R2 &   0.001 &  &  &   0.001 & & \\
Num. obs.  & 5245 & &  & 5245 & &   \\
\bottomrule
\end{tabular}

% \begin{comment}
% 
% \midrule
% \textbf{Intercept}                                        &      29.6058  &        0.671     &    44.098  &         0.000        &        28.290    30.922       \\
% \textbf{C(gender, Treatment(reference='male'))[T.female]} &       3.0417  &        1.684     &     1.807  &         0.071        &        -0.259     6.342       \\
% 
% 
% \begin{tabular}{l|ccc|ccc}
% \toprule
%     & \multicolumn{3}{c|}{Num Regions} & \multicolumn{3}{c|}{Num weeks} \\
%     & $\beta$ & std. err. & p &  $\beta$ & std. err. & p  \\
% \midrule
% Intercept   &   1.5090  &        0.083 &  $^{***}$  &   13.1442  &        0.463   &   $^{***}$   \\
% C(gender)[T.female]  & 0.4179  &        0.209   &  $^{*}$  &  2.7503  &        1.161 & $^{**}$ \\
% \hline
% R2 &   0.001 &  &  &   0.001 & & \\
% Num. obs.  & 5245 & &  & 5245 & &   \\
% \bottomrule
% \end{tabular}
% 
% 
% \end{comment}
\end{table}

\scriptsize

\begin{table}[tb]
\centering
\scriptsize
\caption{\textbf{Topical Bias:} Proportion of the top 200 most associated words (PMI) to each gender that fall into a given category, separated by pre- and post- birth dates of biographies. 
One can see that women tend to have more words related to family, gender and relationships than men. 
However, only pre-1900 the distribution is significantly different ($\chi^2 = 14.33$, $p < 0.01$). On the post-1900 dataset the chi-square test is not significant ($\chi^2 = 5.43$, $p = 0.14$).
}
\begin{tabular}{lcccc}
\toprule
0 -- 1900 & Family & Gender & Relationship & Other  \\                       
\midrule
Men & 0.5 & 1.5 & 0 & 98 \\
Women & 5.0 & 7 & 3 & 85 \\
\midrule
1900 -- Present & Family & Gender & Relationship & Other  \\                       
\midrule
Men & 0.5 & 2.5 & 0 & 97 \\
Women & 3 & 4.5 & 2 & 90.5 \\
\bottomrule
\end{tabular}
\label{table:lexical_categories}
\end{table}

\begin{table}[tb]
\scriptsize
\caption{\textbf{Linguistic Bias:} Comparison of the ratios of abstract terms among positive and negative terms for men and women. Slightly more abstract terms
are used for positive aspects in men's biographies, while slightly more abstract terms are used for
negative aspects in women's biographies. $^{***}: p < 0.001$, $^{**}: p < 0.01$}
\centering
\begin{tabulary}{\linewidth}{lRRLRR}
\toprule
{} &  \% in Men &  \% in Women &    $\chi^2$ &    $w$ &  \% change \\

\midrule
abstract pos       &  27.96 &           25.53 &    933.7*** & 0.04 & 8.69 \\
abstract neg   &          13.47 &        13.69 &   6.26** & 0.005 & -1.62 \\
\bottomrule
\end{tabulary}

\label{table:lib-ratios}
\end{table}

\begin{table}
\scriptsize
\caption{\textbf{Linguistic Bias}: Regression results for the ratio of abstract words among positive and negative words as a function of gender,
profession, and birth century. Women's biographies tend to contain more abstract terms for negativity and less abstract terms for positivity. 
$^{***}: p < 0.001$, $^{**}: p < 0.01$, $^*: p < 0.05$}
\begin{center}
\begin{tabular}{l c c }
\toprule
                 & Abstract positive & Abstract negative \\
\midrule
(Intercept)      & $0.63 \; (0.05)^{***}$  & $0.25 \; (0.05)^{***}$  \\
G[female]        & $-0.02 \; (0.00)^{***}$ & $0.01 \; (0.00)^{**}$   \\
cArchitect       & $0.07 \; (0.05)$        & $0.06 \; (0.05)$        \\
cArtist          & $0.01 \; (0.04)$        & $0.07 \; (0.05)$        \\
cAstronaut       & $-0.04 \; (0.06)$       & $0.00 \; (0.06)$        \\
cAthlete         & $0.03 \; (0.04)$        & $0.05 \; (0.05)$        \\
cBeautyQueen     & $-0.02 \; (0.05)$       & $-0.06 \; (0.05)$       \\
cBusinessPerson  & $0.00 \; (0.09)$        & $-0.02 \; (0.09)$       \\
cChef            & $0.03 \; (0.06)$        & $0.01 \; (0.06)$        \\
cCleric          & $-0.10 \; (0.04)^{*}$   & $0.07 \; (0.05)$        \\
cCoach           & $-0.04 \; (0.04)$       & $0.15 \; (0.05)^{**}$   \\
cCriminal        & $-0.09 \; (0.04)^{*}$   & $0.09 \; (0.05)$        \\
cEconomist       & $-0.01 \; (0.05)$       & $0.15 \; (0.05)^{**}$   \\
cEngineer        & $0.01 \; (0.05)$        & $0.08 \; (0.05)$        \\
cHistorian       & $-0.00 \; (0.07)$       & $0.10 \; (0.07)$        \\
cHorseTrainer    & $-0.06 \; (0.05)$       & $0.03 \; (0.06)$        \\
cJournalist      & $-0.03 \; (0.06)$       & $0.15 \; (0.06)^{*}$    \\
cJudge           & $-0.17 \; (0.04)^{***}$ & $0.02 \; (0.05)$        \\
cMilitaryPerson  & $-0.05 \; (0.04)$       & $-0.02 \; (0.05)$       \\
cModel           & $-0.03 \; (0.05)$       & $0.02 \; (0.06)$        \\
cMonarch         & $-0.07 \; (0.05)$       & $0.01 \; (0.06)$        \\
cNoble           & $-0.06 \; (0.05)$       & $0.03 \; (0.05)$        \\
cOfficeHolder    & $-0.06 \; (0.04)$       & $0.04 \; (0.05)$        \\
cPerson          & $-0.02 \; (0.04)$       & $0.07 \; (0.05)$        \\
cPhilosopher     & $0.05 \; (0.05)$        & $0.11 \; (0.05)^{*}$    \\
cPlayboyPlaymate & $-0.06 \; (0.10)$       & $-0.03 \; (0.10)$       \\
cPolitician      & $-0.06 \; (0.04)$       & $0.05 \; (0.05)$        \\
cPresenter       & $-0.05 \; (0.06)$       & $0.06 \; (0.06)$        \\
cReligious       & $0.04 \; (0.06)$        & $0.11 \; (0.06)$        \\
cRoyalty         & $-0.07 \; (0.04)$       & $0.04 \; (0.05)$        \\
cScientist       & $0.05 \; (0.04)$        & $0.10 \; (0.05)^{*}$    \\
cSportsManager   & $0.01 \; (0.04)$        & $0.06 \; (0.05)$        \\
cent             & $-0.02 \; (0.00)^{***}$ & $-0.01 \; (0.00)^{***}$ \\
\midrule
AIC              & -20917.94               & -21900.42               \\
Num. obs.        & 50965                   & 48942                   \\
\bottomrule
\end{tabular}

\label{table:lib-regression}
\end{center}
\end{table}

\begin{table}[tb]
 \scriptsize
 \caption{\textbf{Meta-data Asymmetries: } Proportion of men and women who have the specified attributes in their infoboxes. Proportions were tested with a chi-square test, with effect size estimated using Cohen's $w$.  
 $^{***}: p < 0.001$, $^{**}: p < 0.01$, $^*: p < 0.05$}% \cite{cohen1992power}
\centering
 % \begin{tabulary}{\linewidth}{lRRLR}
% \toprule
% {} &  \% Men &  \% Women &    $\chi^2$ &    $w$ \\
% \midrule
% birthName       &          4.01 &           11.46 &    4.84* & 0.81 \\
% careerStation   &          8.95 &            1.13 &   6.84** & 0.94 \\
% deathDate       &         32.82 &           19.35 &    5.53* & 0.64 \\
% deathYear       &         44.68 &           25.45 &   8.28** & 0.66 \\
% formerTeam      &          4.40 &            0.24 &    3.94* & 0.97 \\
% numberOfMatches &          8.60 &            1.06 &    6.61* & 0.94 \\
% occupation      &         12.52 &           23.28 &    4.97* & 0.68 \\
% position        &         13.62 &            1.68 &  10.46** & 0.94 \\
% spouse          &          1.56 &            6.86 &    4.10* & 0.88 \\
% team            &         14.06 &            1.97 &  10.39** & 0.93 \\
% title           &          9.17 &           19.65 &    5.59* & 0.73 \\
% years           &          8.95 &            1.12 &   6.84** & 0.94 \\
% \bottomrule
% \end{tabulary}

\begin{tabulary}{\linewidth}{l|RRLR|RRLR}
\toprule
   & \multicolumn{4}{c|}{0 -- 1899} & \multicolumn{4}{c}{1900 -- Present} \\
{} &  \% Men &  \% Women &      $\chi^2$ &     w &  \% Men &  \% Women &     $\chi^2$ &    w \\
\midrule
activeYearsEndDate   &          1.68 &            0.11 &   23.25*** &  3.84 &          2.94 &            1.67 &      0.97 &   -- \\
activeYearsStartYear &          0.64 &            1.08 &       0.31 &    -- &          8.07 &           12.92 &      2.91 &   -- \\
birthName            &          0.53 &            1.02 &       0.44 &    -- &          2.86 &            8.45 &  10.93*** & 1.40 \\
careerStation        &            -- &              -- &        -- &    -- &          8.35 &            1.08 &  48.81*** & 2.59 \\
deathDate            &         15.25 &            7.10 &     9.37** &  1.07 &         12.50 &            9.27 &      1.13 &   -- \\
deathYear            &         16.15 &            7.51 &     9.94** &  1.07 &         13.09 &            9.58 &      1.29 &   -- \\
homepage             &          0.03 &            0.02 &          0 &    -- &          2.92 &            6.43 &     4.22* & 1.10 \\
numberOfMatches      &            -- &              -- &        -- &    -- &          8.06 &            1.02 &  48.58*** & 2.63 \\
occupation           &          1.68 &            1.43 &       0.04 &    -- &          7.51 &           15.69 &    8.90** & 1.04 \\
position             &          0.61 &               0 &  513.34*** & 29.04 &         12.54 &            1.63 &  73.10*** & 2.59 \\
spouse               &          0.44 &            1.51 &       2.57 &    -- &          0.74 &            3.47 &   10.12** & 1.92 \\
team                 &            -- &              -- &        -- &    -- &         12.74 &            1.78 &  67.59*** & 2.48 \\
title                &          1.44 &            1.91 &       0.15 &    -- &          4.94 &           12.49 &  11.53*** & 1.24 \\
years                &            -- &              -- &        -- &    -- &          8.34 &            1.08 &  48.82*** & 2.59 \\
\bottomrule
\end{tabulary}

\label{table:biography-meta-data}
\end{table}

\begin{table}[tb]
\scriptsize
\caption{\textbf{Hyperlink Network Asymmetries:} Comparison of the empirical network and the null models. \textit{M} refers to men and \textit{W} to women. Number of nodes in all networks are 109,529 (0 -- 1899) and 323,762 (1900 -- Present). One can see that in both empirical networks the articles about women link more to other women than we would expect from the null models.}
\centering
\begin{tabulary}{\linewidth}{l|RR|RRRR|RRRR}
% \toprule
% {} &       Clust. Coeff. & Edges (M~to~M) & Edges (M~to~W) & $\chi^{2}$ (M~to~W) & Edges (W~to~M) & Edges (W~to~W) & $\chi^{2}$ (W~to~W)  \\
% \midrule
% Observed        &           0.16 &        90.05\% &         9.95\% &                2.38 &        62.19\% &        37.81\% &            37.83***  \\
% Small World      &           0.16 &        84.45\% &        15.55\% &                0.00 &        84.15\% &        15.85\% &                0.01  \\
% Random          &            0.00 &        84.41\% &        15.59\% &                0.00 &        84.39\% &        15.61\% &                0.00  \\
% In Deg. Seq.    &            0.00 &        85.36\% &        14.64\% &                0.06 &        85.27\% &        14.73\% &                0.05  \\
% Out Deg. Seq.  &          0.00 &        84.43\% &        15.57\% &                0.00 &        84.37\% &        15.63\% &                0.00 \\
% Full Deg. Seq. &          0.00 &        85.34\% &        14.66\% &                0.06 &        85.39\% &        14.61\% &                0.06  \\
% \bottomrule
\toprule
{0 -- 1900} &   Edges &  Clust. Coeff. & Edges (M~to~M) & Edges (M~to~W) & $\chi^{2}$ (M~to~W) & Edges (W~to~M) & Edges (W~to~W) & $\chi^{2}$ (W~to~W)  \\
\midrule
Observed     &  584,879 &           0.16 &        93.10\% &         6.90\% &                0.20 &        69.47\% &        30.53\% &            67.25*** \\
Random    &  415,145 &           0.00 &        92.26\% &         7.74\% &                0.02 &        92.28\% &         7.72\% &                0.02 \\
Small World      &  219,058 &           0.16 &        91.89\% &         8.11\% &                0.00 &        91.53\% &         8.47\% &                0.02 \\
Degree Sequence &  584,879 &           0.00 &        90.22\% &         9.78\% &                0.37 &        90.25\% &         9.75\% &                0.35 \\
\midrule
{1900 -- Present} &    Edges &  Clust. Coeff. & Edges (M~to~M) & Edges (M~to~W) & $\chi^{2}$ (M~to~W) & Edges (W~to~M) & Edges (W~to~W) & $\chi^{2}$ (W~to~W)  \\
\midrule
Observed     &  1,772,793 &           0.11 &        89.47\% &        10.53\% &                3.37 &        54.91\% &        45.09\% &            52.67*** \\
Random    &  1,052,299 &           0.00 &        83.15\% &        16.85\% &                0.03 &        83.21\% &        16.79\% &                0.04 \\
Small World      &   647,524 &           0.11 &        82.51\% &        17.49\% &                0.00 &        82.48\% &        17.52\% &                0.00 \\
Degree Sequence &  1,772,793 &           0.00 &        83.00\% &        17.00\% &                0.02 &        83.11\% &        16.89\% &                0.03 \\
\bottomrule
\end{tabulary}
%&  693,843 
\label{table:biography-network-properties}
\end{table}

\printbibliography

\end{document}